# Origins of chalcogenide perovskite instability

Adelina Carr[a], Talia Glinberg[b], Nathan Stull[c], James R. Neilson[c,d], Christopher J. Bartel[a*]

[a] University of Minnesota, Department of Chemical Engineering and Materials Science, Minneapolis, MN 55455

[b] University of Minnesota, Department of Chemistry, Minneapolis, MN 55455

[c] Colorado State University, Department of Chemistry, Fort Collins, CO 80523

[d] Colorado State University, School of Materials Science & Engineering, Fort Collins, CO 80523

* correspondence to cbartel@umn.edu

## Abstract

Chalcogenide perovskites, particularly II-IV $ABS_3$ compounds, are a promising class of materials for optoelectronic applications. However, these materials frequently exhibit instability in two respects: 1) a preference for structures containing one-dimensional edge- or face-sharing octahedral networks instead of the three-dimensional corner-sharing perovskite framework (polymorphic instability), and (2) a tendency to decompose into competing compositions (hull instability). We evaluate the stability of 81 $ABS_3$ compounds using Density Functional Theory, finding that only $BaZrS_3$ and $BaHfS_3$ are both polymorphically and hull stable, with the $NH_4CdCl_3$-type structure being the preferred polymorph for 77% of these compounds. Comparison with existing tolerance factor models demonstrates that these approaches work well for known perovskites but overpredict stability for compositions without published experimental results. Polymorphic stability analysis reveals that perovskite structures are stabilized by strong *B*-S bonding interactions, while needle structures exhibit minimal *B*-S covalency, suggesting that electrostatic rather than covalent interactions drive the preference for edge-sharing motifs. Hull stability analysis comparing $ABS_3$ to $ABO_3$ analogues reveals a weaker inductive effect in sulfides as a possible explanation for the scarcity of sulfides compared with oxides. The relative instability of $ABS_3$ compounds is further supported by experimental synthesis attempts. These findings provide fundamental insights into the origins of instability in chalcogenide perovskites and highlight the challenges in expanding this promising materials class beyond the few materials that have been reported to date.

## Introduction

Chalcogenide perovskites have garnered significant attention for optoelectronic applications.[1] These materials exhibit smaller bandgaps than oxides,[2] greater thermal and chemical stability than hybrid organic-inorganic halides,[3,4] and additional features like high absorption coefficients and defect-tolerant carrier transport that enhance their functionality.[5,6]

This class of materials follows the general formula $ABX_3$, where $A$ and $B$ are cations and $X$ is a chalcogenide anion (S, Se, Te). This work focuses on the most well-studied subset of these materials where $A$ is divalent, $B$ is tetravalent, and $X$ is sulfur, excluding compounds that are radioactive or contain *f*-electrons due to their limited relevance for optoelectronic applications. While the ideal perovskite structure is cubic, chalcogenide perovskites rarely adopt this high-symmetry structure.[7] Instead, they typically crystallize in distorted perovskite polymorphs due to the larger, more polarizable chalcogenide anions and non-ideal cation-anion radii ratios. Among these, the $GdFeO_3$-type orthorhombic perovskite is the most commonly observed.[1] As shown in **Figure 1a**, this structure consists of a three-dimensional network of tilted corner-sharing $BS_6$ octahedra, with the larger $A$-site cations occupying the interstitial voids. This 3D network plays a crucial role in enabling high and isotropic carrier mobility, direct bandgaps, and other optoelectronic properties.[8] While most research into $ABS_3$ materials focuses on the perovskite structure, these compounds often exhibit a tendency to form alternative structures featuring face- or edge-sharing octahedra, which can lead to different electronic and structural properties.[1,9–11]

The three most common polymorphs of $ABS_3$ chalcogenides (including orthorhombic perovskite) are depicted in **Figure 1**. The $BaNiO_3$-type hexagonal structure (**Figure 1b**) consists of a one-dimensional network of face-sharing octahedra, while the $NH_4CdCl_3$-type “needle-like” structure (**Figure 1c**) is characterized by a one-dimensional framework of edge-sharing octahedra. These three polymorphs are the focus of this study, but it is worth noting that other structures have been experimentally observed, such as “misfit layered structures” and variants containing mixed edge- and face-sharing octahedra.[12–16]

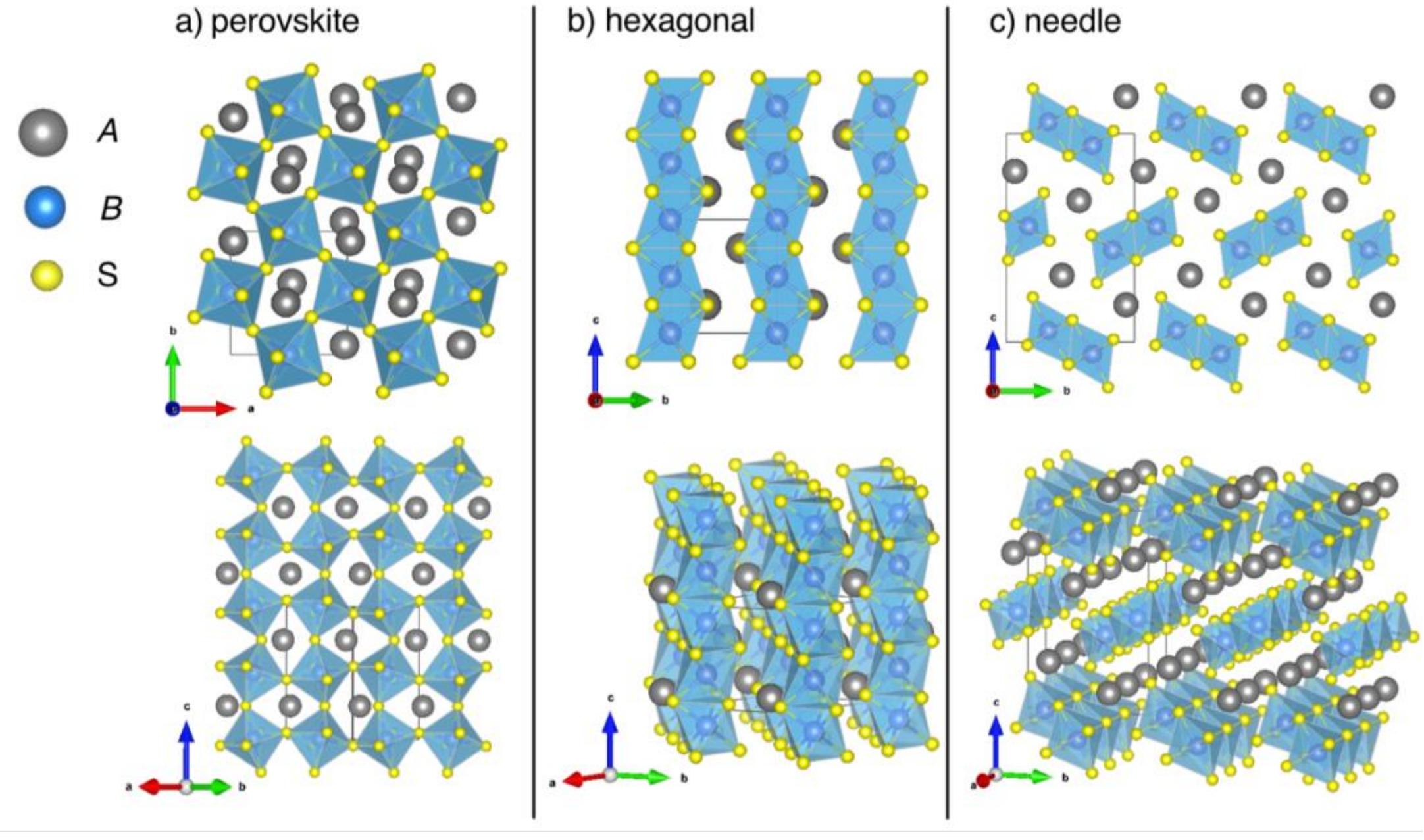

**Figure 1.** Common polymorphs of $ABS_3$ compounds. a) $GdFeO_3$-type orthorhombic perovskite, b) $BaNiO_3$-type hexagonal and c) $NH_4CdCl_3$-type "needle-like" structure. Gray spheres represent *A*-site cations, blue spheres represent *B*-site cations, and yellow spheres represent sulfur.

Although numerous $ABS_3$ compounds within our scope have been synthesized in various structures, only the six compounds spanning *A* = Ba, Sr, Ca and *B* = Zr, Hf have been synthesized in the perovskite structure.[1] $SrZrS_3$ is unique among this group in that it has been synthesized in both the perovskite and needle structures, with perovskite emerging as the higher temperature phase.[17,18] $BaZrS_3$ is the most well studied for its promising optoelectronic properties and has been synthesized at a number of conditions, including low-temperatures colloidal synthesis.[5,19–23] However, the most common synthetic approach to realize these materials is high-temperature sulfurization of their oxide perovskite analogues,[19] with $CaHfS_3$ and $CaZrS_3$ requiring comparatively higher temperatures than the *A* = Ba, Sr counterparts.[21,24]

Previous computational studies have evaluated the thermodynamic stability of various subsets of chalcogenide perovskites.[12,25–29] For example, Huo, *et al*. used Density Functional Theory (DFT) with the PBE functional[30] to study 168 $ABX_3$ compounds (Including *X* = O, S, Se) in four different crystal structures: cubic perovskite, orthorhombic perovskite, hexagonal, and needle. They found that ~30% adopted a perovskite ground state, including sulfides with *A* = Ba, Ca and *B* = Hf, Zr.[28] Brehm *et al*. used LDA[31] to study $ABS_3$ compounds with $d^0$ configurations[12] (including I-V and II-IV valences for *A*-*B*) and found none had a perovskite ground state; instead, $BaZrS_3$ was calculated in that study to prefer the needle structure. Their finite-temperature analysis, however, showed that perovskites are stabilized at elevated temperatures for systems with *A* = Ba, Sr, Sn, Pb and *B* = Zr. In addition to polymorphic stability, some studies have also considered the stability of $ABX_3$ compounds against decomposition into competing phases.[25,26] For example, Körbel *et al*. computed ternary phase diagrams for a broad range of $ABX_3$ compounds (not limited to chalcogenides), identifying $BaHfS_3$, $BaZrS_3$, and $BaZrSe_3$ as the only chalcogenide perovskites within 25 meV/atom of stability according to their calculated convex hull.[25]

In parallel to thermodynamic analyses with DFT, tolerance factor models have long been used to assess the tendency for $ABX_3$ materials to crystallize in perovskite or non-perovskite structures (in this work, perovskite refers to materials comprised of corner-sharing $BX_6$ octahedra surrounding an *A*-site cation). The most well-known of these models is the Goldschmidt tolerance factor,[32] which relies only on geometric considerations and the ionic radii of constituent elements, usually Shannon's ionic radii calibrated for oxides.[33] While this model and subsequent refinements[34] have worked well for oxide and halide perovskites, its accuracy declines when applied to non-oxide chalcogenides.[1,35,36]

Recent efforts have sought to improve tolerance factor models for chalcogenide perovskites. In 2022, Jess *et al.* introduced a modified tolerance factor, recognizing that the significantly lower electronegativity of sulfur compared to oxygen results in greater covalent bonding.[35] To account for this, their model introduced an electronegativity-weighted correction to the Goldschmidt tolerance factor:

$$t^* = \frac{\frac{\Delta\chi_{(A-X)}}{\Delta\chi_{(A-O)}}(r_A + r_X)}{\sqrt{2}\frac{\Delta\chi_{(B-X)}}{\Delta\chi_{(B-O)}}(r_B + r_X)} \qquad (1)$$

Where $\Delta\chi$ represents the electronegativity differences between the indicated sites, and $r_A$, $r_B$ and $r_X$ are the ionic radii for the *A*, *B*, and *X* sites, respectively. This electronegativity-modified tolerance factor successfully grouped 9 experimentally known chalcogenide perovskites within $0.94 < t^* < 1.12$ ($BaZrS_3$, $BaHfS_3$, $SrZrS_3$, $SrHfS_3$, $CaZrS_3$, $CaHfS_3$, $EuZrS_3$, $EuHfS_3$, and $BaUS_3$) with no known non-perovskites falling within this interval.

In 2024, Turnley *et al.* introduced another tolerance factor based approach,[36] noting a key limitation in previous models: the ionic radii data used in traditional tolerance factors (Shannon's dataset) was empirically derived from metal oxides and fluorides, assuming ionic bonding. However, the lower electronegativity of sulfides yields a tendency for stronger covalent interactions, making the oxide-calibrated radii less applicable to chalcogenide perovskites. Shannon recognized this issue and created a separate dataset of metal-sulfide-derived radii, although this dataset remains less extensive.[37] Turnley *et al.* used this sulfide-derived radii dataset (extrapolated where necessary) to establish a refined perovskite stability criterion. Their screening method involves three metrics. First, materials were screened using the octahedral factor ($\mu = r_B/r_X$) to determine whether *B* and *X*-site ions were appropriately sized to form stable octahedra. Materials passing the octahedral factor test ($0.414 < \mu < 0.732$) were then evaluated based on a modified Goldschmidt tolerance factor using sulfide-derived radii ($t^S$), constrained within $0.865 \leq t^S \leq 0.965$ along with an electronegativity-based metric, $\chi_{diff} = 1/5\ (3\chi_S - \chi_A - \chi_B)$, requiring $\chi_{diff} \geq 1.025$ for perovskite stability. In their study, Turnley *et al.* applied this multi-step screening method to 100 compounds from the Materials Project database of DFT calculations,[38] 24 of which are experimentally known chalcogenide perovskites. Among those, their tolerance factor model correctly identified 18 as lying within the proposed stability window, including *f*-electron and radioactive elements, as well as materials belonging to different valence families (e.g., III–III compounds). When restricting the comparison to the six experimentally reported chalcogenide perovskites within our scope, the Turnley model successfully places all six within the predicted perovskite stability window.

While tolerance factors are a valuable tool for assessing perovskite stability, they intrinsically focus only on polymorphic stability and do not probe the tendency for decomposition into competing compositions (hull stability).[39] Both polymorphic and hull stability at 0 K can be evaluated using DFT calculations, though they involve slightly different energetic comparisons. Polymorphic stability is determined by comparing total energies across different polymorphs at fixed chemical composition, with the lowest-energy structure defined as the ground state. The energy difference between a given polymorph and the ground state is denoted as $\Delta E_{gs}$. Hull stability, assessed via the convex hull formalism, requires computing formation energies for all ground-state structures in a given chemical space (e.g., *A*S, $BS_2$, etc., for *A-B-S* in $ABS_3$). A material is considered hull-stable if its formation energy is negative and lower than any linear combination of competing phases in that chemical space. The decomposition energy ($\Delta E_d$) quantifies the energy difference relative to the most stable phase-separated state, with stability requiring $\Delta E_d \leq 0$. Materials with $\Delta E_d > 0$ are computed to be unstable with respect to decomposition but may still be synthetically accessible metastable phases if $\Delta E_d$ and $\Delta E_{gs}$ are not too large.[40] Such metastable phases remain of high interest. For example, $SrZrS_3$ is thermodynamically stable in the needle phase, but its perovskite polymorph has been synthesized under high-temperature conditions.[17,18] Likewise, $BaZrSe_3$ and alloyed variants have been successfully synthesized and retained in the perovskite structure[41,42] despite their thermodynamic instability.[29] In the Results and Discussion section, we examine $\Delta E_{gs}$ and $\Delta E_d$ of previously synthesized metastable materials in the context of prior studies[40,43] that have proposed heuristic thresholds and statistical trends relating stability and synthesizability.

In this study, we used DFT calculations at the meta-GGA level of theory[44] to assess the polymorphic and hull stability of 81 $ABS_3$ chalcogenides and probed the two recently introduced tolerance factors in the context of our DFT-calculated thermodynamics. Our study highlights the sparsity of thermodynamically stable chalcogenide perovskites and proposes thermodynamic and chemical origins for their instability.

## Methods

**Computational Details.** All DFT calculations were performed using the Vienna Ab Initio Simulation Package[45,46] (VASP, version 6.4.1) and the projector augmented wave (PAW) method.[47,48] A plane-wave energy cutoff of 520 eV, and Γ-centered k-point grid with a minimum spacing between k-points of 0.22 $Å^{-1}$ (KSPACING in VASP), were used for all calculations. The convergence criteria were set as $10^{-6}$ eV for electronic optimization and 0.03 $eV \cdot Å^{-1}$ for ionic relaxation. To assess the thermodynamic stability of various $ABS_3$ compounds, the r²SCAN meta-generalized gradient approximation (meta-GGA) density functional was applied.[44] Although r²SCAN offers only modest improvements over GGA for decomposition energies, it significantly improves formation energy accuracy—often reducing errors by a factor of 1.5 to 3.[49] This makes it particularly well-suited for our polymorphic stability analysis, which depends on precise total energy differences. These improvements have been demonstrated across transition metal compounds, main group compounds, and chalcogenides.

The initial structures for the 81 $ABS_3$ compounds investigated in this study were generated by applying atomic substitutions to three prototype structures queried from the Materials Project database[38]: $GdFeO_3$ (orthorhombic perovskite), $BaNiO_3$ (hexagonal), and $SrZrS_3$ (needle-like), which are representative of stable polymorphs in their respective structure types. After substitution, all structures were optimized using the calculation settings mentioned above. Polymorphic stability was assessed by comparing DFT total energies at 0 K across the three polymorphs for each composition. Hull stability was evaluated by querying all competing phases in each *A*-*B*-S phase diagram from the Materials Project database[38] and recomputing all formation energies using the aforementioned calculation settings. As our analysis is restricted to 0 K ground-state energetics, it does not capture phases that may become thermodynamically stable at elevated temperatures or pressures.

The LOBSTER package[50] was used for COHP[51] analysis. Charge analysis was performed via the Bader method.[52,53] Structures and charge densities were analyzed using VESTA[54]. Pymatgen was used to prepare and analyze DFT and LOBSTER calculations.[55,56]

For tolerance factor analysis, ionic radii were taken directly from the supplementary data of the respective studies (Jess *et al.*, Turnley *et al.*) and used as reported[35,36] with the exception of $Pb^{4+}$ and $Si^{4+}$, whose sulfide-derived radii were not reported in the Turnley dataset. For $Pb^{4+}$, the radius was extrapolated from available coordination states. For $Si^{4+}$, the oxide-derived radius was used instead.

**Experimental details.** Reactions targeting the synthesis of $CaTiS_3$ and $BaZrS_3$ were carried out by the attempted boron-assisted sulfidation of the corresponding oxide perovskites, $CaTiO_3$ and $BaZrO_3$.[57,58] All manipulations were performed in air unless explicitly specified. The oxide precursors were prepared by mixing stoichiometric amounts of the alkaline earth carbonates, $CaCO_3$ (Sigma-Aldrich 99%) and $BaCO_3$ (J.T. baker 99%), with the corresponding transition metal oxides, $ZrO_2$ (Sigma-Aldrich 99%) and $TiO_2$ (Aldrich Chemical 99.9%), in an agate mortar with a small amount of isopropanol for 15 min and evaporated to dryness. The powders were then pelletized in a 0.5 in die to 2 tons of force, placed in an alumina crucible and heated in a tube furnace to 1100 °C at a ramp rate of 10 °C/min and held at temperature in air for 24 h. For the sulfidation reaction, stoichiometric amounts of oxide perovskite precursor and sulfur (NOAH Technology Corporation 99.5%), with an excess 2.2 molar ratio of boron (Aldrich Chemistry 95%) targeting $AMO_3$ + 2.2 B + 3 S, were ground in an agate mortar for 15 min and loosely packed within an alumina crucible that was held in an extruded quartz ampule. These ampules (16 mm O.D.; 14 mm I.D.) were flame sealed under vacuum (~15 mTorr) and placed into a muffle furnace. The reaction with $CaTiO_3$ initially showed significant reaction with the $SiO_2$ ampule. As such, the reaction was repeated in a stainless-steel tube with 3/8 in inner diameter (no $Al_2O_3$ crucible). The tube was crimpled closed and then welded using an MRF Arc Melt Furnace under ~0.85 atm of argon. All samples were then heated at a ramp rate of 5 °C/min to 300 °C and 600 °C with holding times of 5 h each, then finally to 1000 °C for a holding time of 36 h. The samples were let to furnace cool and opened in air for analysis.

Powder X-ray diffraction (PXRD) measurements were conducted using a Cu Kα X-ray source and a LynxEye XE-T position-sensitive detector on a Bruker D8 Discover diffractometer to verify phase purity. Before measurement, the samples were placed on a "zero diffraction" Si wafer and immobilized with petroleum jelly. Quantitative phase analysis employing the Rietveld method was performed using TOPAS v6.

**Results and discussion**

To better understand previous experimental and computational results, we considered a subset of the materials evaluated by Jess *et al.*[35] We selected 6 $ABS_3$ compounds predicted by this tolerance factor to be perovskite as well as 39 $ABS_3$ compounds that lie near the borders of perovskite stability with respect to this tolerance factor. An additional 36 $ABS_3$ compounds were computed to consider all $A/B$ combinations spanned by these 45 compounds. For each of these 81 compounds, we performed DFT calculations in the distorted perovskite, needle, and hexagonal polymorphs using the r$^2$SCAN meta-GGA functional.[44] In **Figure 2**, we show the DFT-predicted thermodynamic stability of the analyzed compounds. These stabilities were assessed in two ways: $\Delta E_{gs}$ indicates the difference in energy between the perovskite structure and the corresponding ground-state polymorph; $\Delta E_d$ indicates the decomposition energy for the ground-state polymorph and is calculated by considering all competing phases available in the Materials Project database[38] (recomputed with r$^2$SCAN in this work). This $\Delta E_d$ should be interpreted as a lower bound, as additional competing phases not currently present in the Materials Project database could further destabilize a given compound. For example, we did not systematically investigate other perovskite-related compounds (e.g., Ruddlesden-Popper) in our hull analysis unless these compounds were already present in Materials Project. All $\Delta E_{gs}$, $\Delta E_d$ and calculated ground-state structures are listed in **Table S1**, along with the predicted decomposition reactions and number of stable compounds in each $A$-$B$-S phase diagram. An $ABS_3$ compound is thermodynamically stable in the perovskite structure if $\Delta E_{gs}$ and $\Delta E_d$ are both $\leq 0$. The magnitude of thermodynamic instability for a perovskite-structured $ABS_3$ is then given by the sum of $\Delta E_{gs}$ and $\Delta E_d$. Compounds that have been previously synthesized in any polymorph are highlighted in blue, and the shape of the marker indicates the DFT-calculated ground-state structure.

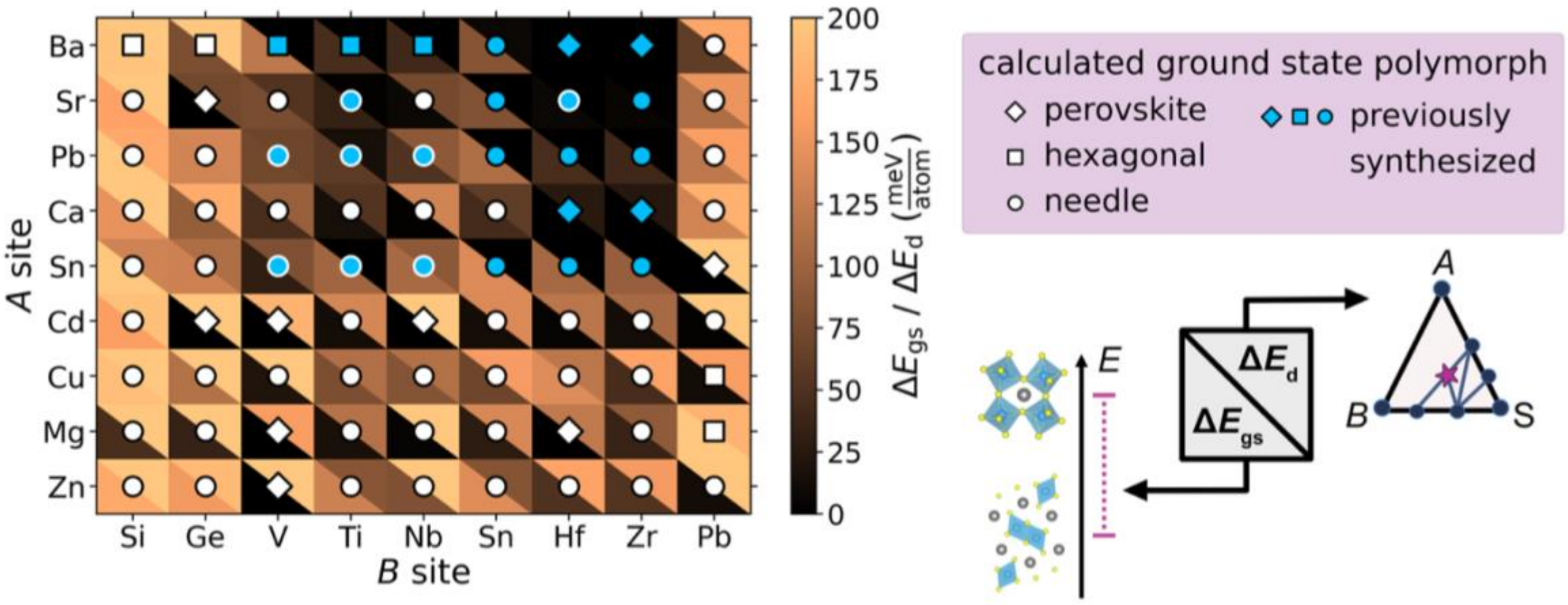

**Figure 2.** Heatmap summarizing the thermodynamic stability of 81 selected $ABS_3$ compounds. Each square corresponds to a unique $ABS_3$ compound, with $A$-site elements on the $y$-axis and $B$-site elements on the $x$-axis. Each lower left triangle shows the polymorphic instability, $\Delta E_{gs}$, for the perovskite structure (a value of 0 indicates that perovskite is the ground-state structure). Each upper right triangle indicates the decomposition energy, $\Delta E_{d}$, relative to the $A$-$B$-S convex hull (a value of 0 indicates that the ground-state structure is thermodynamically stable with respect to decomposition into competing compounds). Marker shapes indicate the DFT-predicted ground-state polymorph, and blue-filled markers denote compounds that have been previously synthesized (in any structure). Markers outlined in white indicate compounds for which the experimentally reported structure does not match the DFT-predicted ground-state structure.

Our calculations show that all 24 of the previously synthesized $ABS_3$ materials (in any polymorph) have $\Delta E_{d} < 90$ meV/atom with 20/24 having $\Delta E_{d} < 30$ meV/atom. We should note that in some cases specialized synthetic approaches, such as high total pressure ($\geq 2$ GPa), were used to access the materials (as is the case with $A$SnS$_3$, where $A$ = Ba, Pb, Sr).[14,59] Among the materials considered that have not yet been synthesized, seven have $\Delta E_{d} \leq 90$ meV/atom, including three with $\Delta E_{d} < 61$ meV/atom ($CaTiS_3$, $CaSnS_3$ and $SrVS_3$). For context, a previous study analyzing over 29,000 materials in the Materials Project database found that 90% of synthesized metastable materials are within 67 meV/atom of the hull.[40] While these numbers provide a helpful reference for assessing metastability, it is important to emphasize that being below this threshold does not guarantee synthesizability. It is possible that there are no thermodynamic conditions under which a specific material is thermodynamically accessible, even if it has a weakly positive $\Delta E_{d}$. These results highlight the sparsity of $ABS_3$ compounds that are thermodynamically stable or weakly unstable with respect to decomposition into competing phases in each $A$-$B$-S chemical space.

As for $\Delta E_{gs}$, our calculations predict that 12 of 81 compounds have perovskite ground states ($\Delta E_{gs} = 0$). This includes 4 of the 6 compounds that have been previously synthesized in the perovskite structure ($BaHfS_3$, $BaZrS_3$, $CaHfS_3$ and $CaZrS_3$). The two remaining experimentally known perovskites within our scope are $SrZrS_3$ and $SrHfS_3$, which have marginal polymorphic instability ($\Delta E_{gs}$ of 6 and 8 meV/atom, respectively). Importantly, many of these compounds with perovskite ground states exhibit hull instabilities (as discussed later). Of $ABS_3$ compounds that have been synthesized in non-perovskite polymorphs, the least unstable perovskite is $SnVS_3$ with $\Delta E_{gs}$ = 31 meV/atom. Of the 24 $ABS_3$ compounds that have been synthesized in any structure, 16 have been synthesized in the DFT-calculated ground state structure. Six of the remaining eight compounds ($ABS_3$ with $A$ = Pb, Sn and $B$ = V, Nb, Ti) have only been reported in the misfit layered structure, which is not included in our calculations. Among the remaining two, one is $SrHfS_3$, where the experimentally observed structure (perovskite) is calculated to be just 8 meV/atom above the calculated ground-state structure (needle). The other is $SrTiS_3$, which has been reported to form in a hexagonal structure or a related hexagonal misfit structure.[14,20,60–62] However, our calculations predict the needle structure as the ground state, with the hexagonal structure significantly higher in energy (81 meV/atom above needle). Interestingly, several other computational studies have also identified the needle structure as the ground state,[12,28,29] a notable discrepancy between theoretical predictions and experimental observations. As with $\Delta E_d$, it is possible that $ABS_3$ compounds with $\Delta E_{gs} > 0$ could be synthesized in the perovskite structure. A previous study proposed using the energy of the amorphous phase as a physically meaningful upper bound for $\Delta E_{gs}$,[43] but this relatively generous upper bound varies widely from compound to compound. In the absence of exotic synthetic approaches, the probability of accessing a perovskite-structured $ABS_3$ should decrease substantially for compounds with larger $\Delta E_{gs}$, and it is notable that the largest $\Delta E_{gs}$ of any previously reported perovskite is only 8 meV/atom ($SrHfS_3$).

Assessing polymorphic and hull stability together, $ABS_3$ compounds where perovskites could plausibly be synthesized are indicated by black or nearly black squares in **Figure 2**. Six of the 81 compounds we calculated have been previously synthesized in the perovskite structure ($BaZrS_3$, $BaHfS_3$, $CaZrS_3$, $CaHfS_3$, $SrZrS_3$, $SrHfS_3$). $BaZrS_3$ and $BaHfS_3$ are the only compounds in our study calculated to be ground-state perovskites ($\Delta E_{gs} = 0$) and stable against competing phases ($\Delta E_d \leq 0$). $CaZrS_3$ and $CaHfS_3$ are also calculated to have perovskite ground states but are slightly above the convex hull ($\Delta E_d$ = 25 and 27 meV/atom, respectively). $SrZrS_3$ and $SrHfS_3$ are calculated to have needle ground states that lie on the convex hull ($\Delta E_{gs}$ = 6 and 8 meV/atom, respectively). Considering all 81 compounds in our study, two key observations emerge: (1) the needle structure is overwhelmingly preferred as the ground-state structure, accounting for 77% of the 81 materials analyzed, and (2) thermodynamically stable perovskites are sparse with only two of 81 compounds having $\Delta E_{gs} = 0$ and $\Delta E_d \leq 0$.

Given the limited number of stable perovskites and the challenges in synthesizing those calculated to be unstable, it is natural to ask whether tolerance factors can reliably predict perovskite formability in chalcogenides beyond the previously synthesized materials used to develop these models. In **Figure 3a**, we show a comparison between the Jess tolerance factor[35] ($x$-axis) and the DFT-calculated $\Delta E_{gs}$ ($y$-axis). The gray-shaded region marks the range where this tolerance factor predicts a stable perovskite ($0.92 < t^* < 1.10$). For clarity, this figure shows a zoomed-in view of the data with the full range available in **Figure S1**. This comparison shows that Jess tolerance factor predictions largely deviate from the DFT predictions for many hypothetical compounds. For example, $SnGeS_3$ strongly favors the needle structure ($\Delta E_{gs}$ = 131 meV/atom) but is predicted to be a stable perovskite by the tolerance factor ($t^* = 0.99$).

A similar comparison is shown in **Figure 3b** for the Turnley tolerance factor,[36] highlighting only the 45 compounds we assessed that pass the octahedral factor filter ($0.414 < \mu < 0.732$). As was done in Ref. [36], each of these compounds are shown with respect to the sulfide-adjusted Goldschmidt tolerance factor $t^S$ ($x$-axis) and the electronegativity difference, $\chi_{diff} = 1/5\ (3\chi_S - \chi_A - \chi_B)$ ($y$-axis). Points are colored by their calculated $\Delta E_{gs}$ with lighter points being less stable in the perovskite structure. Turnley *et al.*, identified a perovskite stability window in the range $0.865 < t^S < 0.965$ and $\chi_{diff} > 1.025$ (shaded gray in **Figure 3b**). Five materials with DFT-predicted perovskite ground-states from our study fall within this window, along with four other compounds calculated to have non-perovskite ground states. There appears to be a relationship between electronegativity and energetic ordering, where compounds with a lower $\Delta E_{gs}$ tend to cluster at higher $\chi_{diff}$ values. There is no clear trend along the $t^S$ axis, reinforcing that this modified Goldschmidt model also fails to capture a trend with energetic ordering. Calculated values for both tolerance factors and their predicted stable structures are provided in **Table S1**, along with the ionic radii, μ, and $\chi_{diff}$ values used in the calculations.

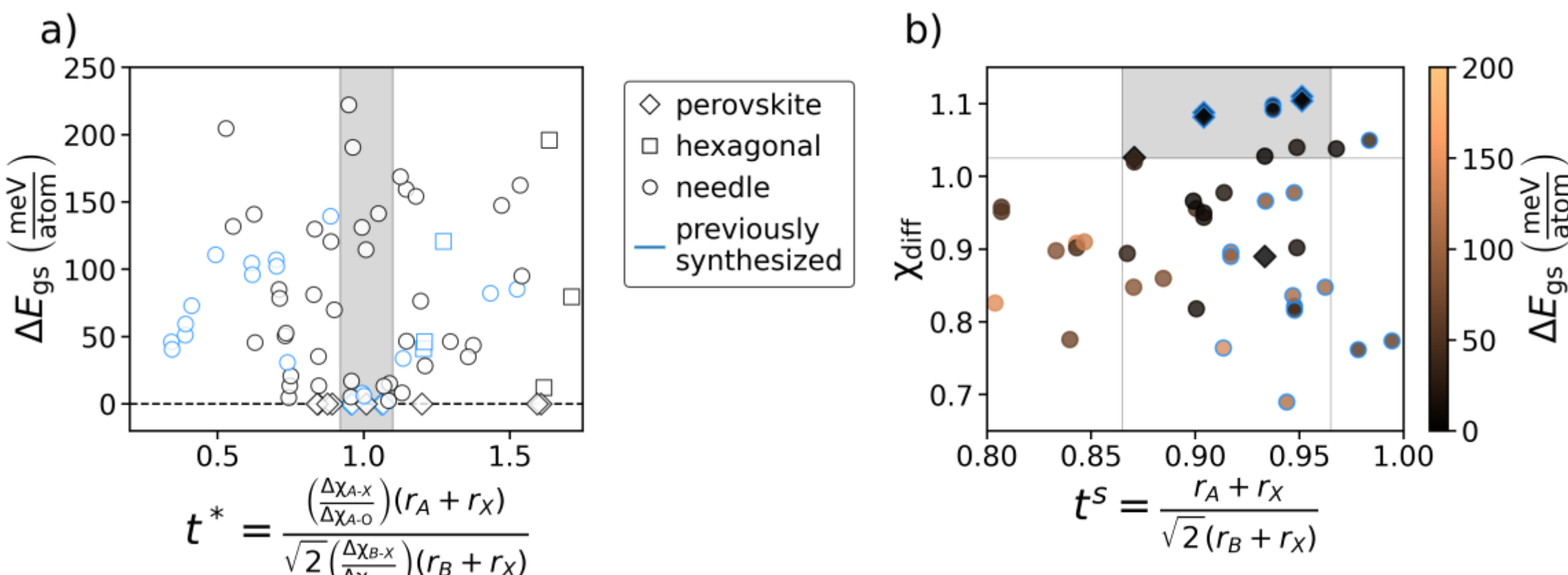

**Figure 3.** a) DFT-predicted energy difference between the perovskite structure and the corresponding ground-state structure ($\Delta E_{gs}$) plotted against the Jess tolerance factor ($t^*$), where points at $\Delta E_{gs} = 0$ indicate a perovskite ground state. The gray region denotes the $t^*$ range where this tolerance factor predicts perovskite stability. b) The electronegativity difference, $\chi_{diff}$, plotted against the Turnley tolerance ($t^S$). The color bar corresponds to the same DFT-calculated $\Delta E_{gs}$. The gray region denotes the $t^S$ range above a certain $\chi_{diff}$ where the Turnley tolerance factor predicts stable perovskites. In both plots, blue-outlined data points denote previously synthesized compounds (in any structure), and marker shapes indicate the DFT-predicted ground-state structure: diamonds for perovskite, circles for needle, and squares for hexagonal.

Together, these analyses highlight the need for a more comprehensive framework to accurately predict chalcogenide perovskite stability. In both cases, there is limited correlation between the tolerance factor and $\Delta E_{gs}$ as we would expect $\Delta E_{gs}$ to be minimal in the gray shaded region and increase as $t^*$ or $t^S$ deviates from the perovskite-stable region. Furthermore, while both models perform well for previously synthesized materials, their boundaries were empirically defined based on these previously synthesized perovskites. Our results suggest that even electronegativity-adjusted tolerance factors often overpredict chalcogenide perovskite stability, emphasizing that the interactions in these systems are more nuanced than those captured by these models. Motivated by this, we aimed to uncover the factors that dictate polymorphic and hull (in)stability.

**Analysis of polymorphic instabilities**

To go beyond geometric considerations and better understand why chalcogenides favor non-perovskite motifs (e.g., edge-sharing octahedra), we turn to the role of covalency and chemical bonding—particularly how the lower electronegativity of sulfur compared to oxygen may contribute to destabilizing the corner-sharing network that is so common in oxides. Previous studies have shown that even in oxide perovskites, increased covalency leads to destabilization of the perovskite, which is further exacerbated by oxide ion polarization.[63] Here, integrated Crystal Orbital Hamilton Population (ICOHP)[51] was used as an analytical tool to observe trends in covalent interactions among $ABS_3$ polymorphs. We mainly compare the perovskite and needle structures due to the strong preference for the needle structure exhibited by the compounds studied in this work.

We illustrate in **Figure 4** how distinct covalent interactions emerge in these two structure types by presenting $CaHfS_3$ as a representative example. In **Figure 4a** and **4c**, we compare the COHP of $CaHfS_3$ in the perovskite (a) and needle structures (c). Both structures exhibit similar S–S antibonding interactions near the Fermi level, arising from the relatively large electron cloud of the sulfide anion, overlapping within and between octahedra in each structure. In contrast, the Hf–S bonding in the perovskite structure is markedly stronger with a large peak in -COHP > 0 (bonding interactions), whereas the needle structure exhibits effectively no Hf-S covalency. In **Figure 4b** and **4d**, we compare the real space charge densities for this same energy interval in each structure. Within the chosen 2D slices of the charge density, the perovskite structure exhibits a linked network of tilted corner-sharing octahedra with substantial covalent Hf-S bonding. In contrast, the edge-sharing octahedra of the needle structure have very little charge density lying between Hf and S ions, instead favoring high density on each S ion.

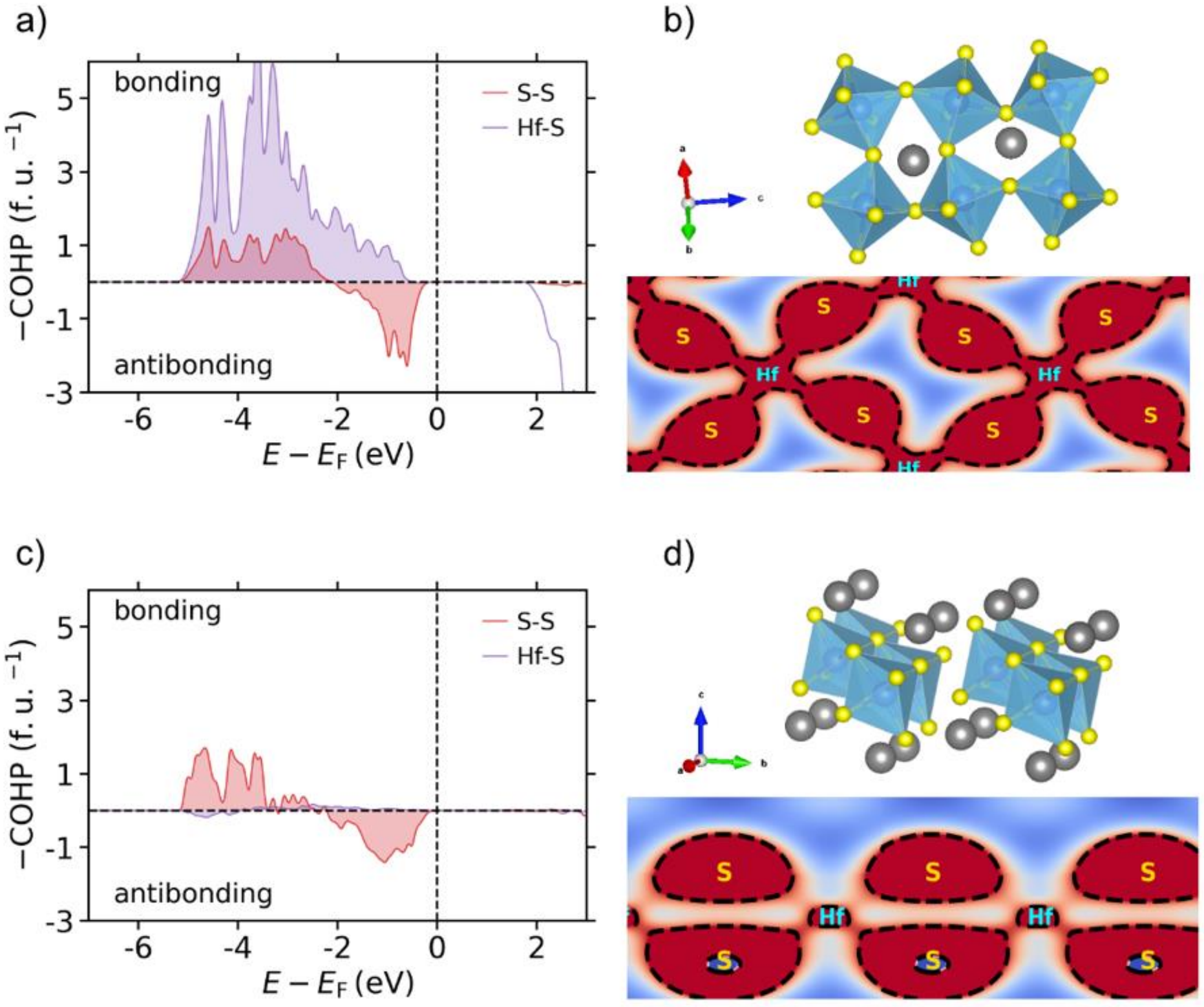

**Figure 4.** COHP analysis and charge density visualization for $CaHfS_3$ in perovskite and needle structures. (a,c) COHP curves plotted with respect to the Fermi energy ($E_F$) for perovskite and needle structures, respectively. Purple curves represent Hf–S interactions; red curves represent S–S interactions. COHP values are normalized per formula unit. (b,d) Two-dimensional slices of the DFT-calculated electron density as viewed along [0 0 1] and [-1 1.9 -2.5] directions for the perovskite and needle structures, respectively. Contours delineate regions of equal density. The needle phase exhibits significantly weaker *B*–S bonding interactions and reduced charge density between *B* and S sites compared to the perovskite structure.

Analysis of our COHP results suggests that covalent interactions tend to preferentially stabilize the perovskite structure over the needle structure, which is surprising given the overwhelming stability of the needle structure. To assess whether there is a broader trend, we compare ICOHPs of needle vs. perovskite across the $ABS_3$ materials within our scope. The bonding trends observed in $CaHfS_3$ generalize across the materials class. As shown in **Figure 5**, covalent *B*–S interactions differ substantially between the two structure types. Compounds that adopt the perovskite ground state (and those with low $\Delta E_{gs}$) typically exhibit strong *B*–S covalency within their $BS_6$ octahedral units. In contrast, compounds that crystallize in the needle structure often show little to no *B*–S covalency, and consistently less than their perovskite counterparts. For clarity, a zoomed-in version of **Figure 5** is provided in **Figure S2** to better distinguish individual scatter points. Due to the relatively large radius of sulfur (compared to oxygen), both structure types display considerable S–S overlap. As shown in **Figure S3**, many compounds exhibit similar S–S antibonding interactions near the Fermi level in both the perovskite and needle structures.

This analysis reveals that the small number of perovskites that are lower in energy than their needle counterparts are stabilized by *B*-S covalent bonding interactions. Recalling that the needle structure is by far the most common ground state among $ABS_3$ compounds (77% of compounds analyzed in this work), our results rule out covalency as the origin of this preference for the needle structure. It is instead plausible that the needle structure offers a more favorable electrostatic arrangement. This notion is supported by previous work from Young and Rondinelli on halide perovskites (e.g., $CsPbI_3$), where electrostatic interactions were found to be the primary origin of stability in the needle structure.[64]

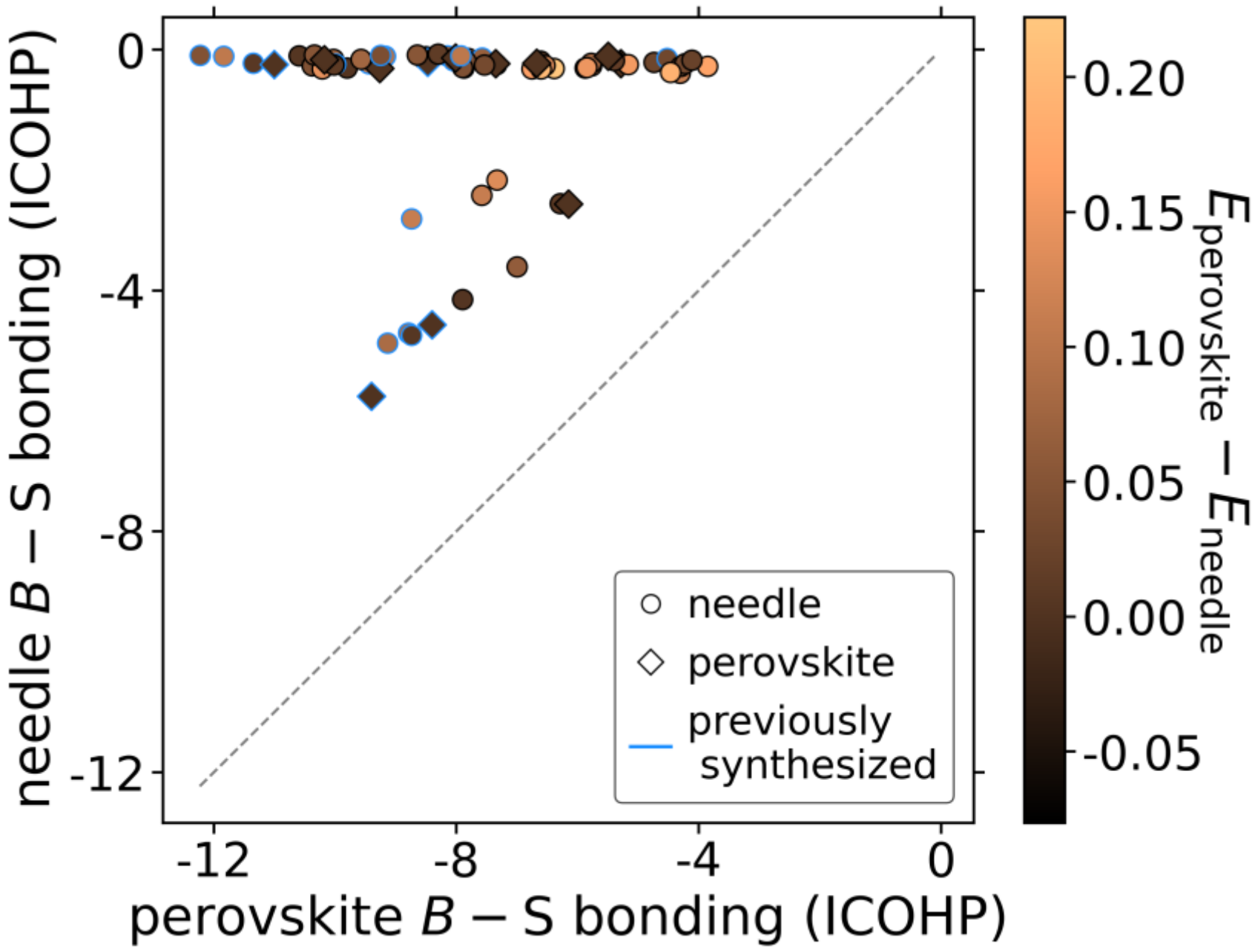

**Figure 5.** Parity plot comparing integrated COHP (ICOHP) values for *B*–S bonding interactions in the needle and perovskite structures, normalized per formula unit and integrated from 0 to –7 eV below the Fermi level. More negative values indicate more bonding interactions. Points are colored according to the difference in energy from the perovskite to the needle structures. Points outlined in blue correspond to previously synthesized materials. Diamond-shaped markers represent materials that have a perovskite ground state, while circles represent a needle ground state (according to our calculations). Because we compare needle and perovskite only, we have excluded materials with a hexagonal ground state. Here, we have only considered materials with differing *A*- and *B*-site atoms. The units for ICOHP are arbitrary.

**Analysis of hull instabilities**

While our COHP analysis sheds light on how covalent interactions influence polymorphic preference, $\Delta E_{gs}$ alone does not provide a complete picture of stability. **Figure 2** revisits this point, illustrating how some materials, despite being polymorphically stable, remain thermodynamically unstable due to their tendency to decompose into competing compounds in a given *A*-*B*-S chemical space. Much of the perovskite literature, particularly studies involving tolerance factors, tends to focus exclusively on polymorphic stability while neglecting stability with respect to competing compounds. This oversight often leads to false positives in theoretical predictions, where materials appear stable in the perovskite structure ($\Delta E_{gs} = 0$) but are thermodynamically unstable due to highly positive decomposition energies (e.g., $CaNbS_3$ in **Figure 2**).

To test the validity of some of these calculations, we performed experimental synthesis reactions of materials predicted to be stable and unstable. An effective method for the synthesis of these materials is to react mixtures of binary oxides or ternary oxides with boron and sulfur.[57] The formation of low-melting and separable $B_2O_3$ (via solution chemistry or via vapor transport) assists in facilitating both the kinetics and thermodynamics of the reaction, as has been shown for refractory actinide sulfides.[65] $BaZrS_3$, predicted to be stable, is easily synthesized by this method (**Figure S3a**). $CaTiS_3$, predicted to lie above the convex hull ($\Delta E_d$ = 52 meV/atom), instead phase separates to CaS, $TiS_3$, and $TiS_2$; we also observe the formation of $CaB_2O_4$ (**Figure S3b**). Our computed decomposition reaction predicts that $CaTiS_3$ should decompose into CaS and $TiS_2$, consistent with the appearance of these two impurity phases after synthesis. The additional formation of $CaB_2O_4$ leads to Ti excess relative to Ca and the additional formation of a $TiS_3$ byproduct, which is calculated to be a stable competing compound in the Ti-S phase diagram. These results emphasize that it is important to understand the origin of the thermodynamic instability of ternary chalcogenides, particularly considering the stability of their oxide analogues (e.g., $CaTiO_3$ vs. $CaTiS_3$).

Ternary oxides are more commonly observed than their sulfide counterparts. For example, > 250 $ABO_3$ materials have been synthesized in the perovskite structure,[34] compared with ~25 $ABS_3$ perovskites (including all possible *A*/*B* sites).[1] To quantify the hull instability of $ABS_3$ compounds (in any structure) compared with their analogous $ABO_3$ compounds, we queried Materials Project[38] for the 81 $ABO_3$ oxides generated from the 81 sulfides studied in this work. Among these, 51 were available in Materials Project, of which 31 were found to lie on the convex hull. In contrast, only 15 of the 81 sulfides from our calculations are hull-stable. **Table S2** lists all available values of $\Delta E_d$ for this analysis. Even for those $ABS_3$ compounds that are thermodynamically stable (on the convex hull), their stability is marginal compared with their $ABO_3$ analogues. In **Figure 6**, we illustrate how the thermodynamic driving force to form ternary sulfides is generally weaker than the driving force to form the analogous oxide by showing 0 K DFT reaction energies for the formation of six $ABX_3$ compounds from $AX + BX_2$ reactants. In all cases, there is a larger driving force for $ABO_3$ compared with $ABS_3$ formation. One might attribute this to the inherent stability of binary sulfides, but this argument alone is insufficient, as oxides also form highly stable binary compounds, usually with higher melting points than their sulfide analogues. This suggests that additional mechanisms must be at play in stabilizing ternary oxides compared with sulfides.

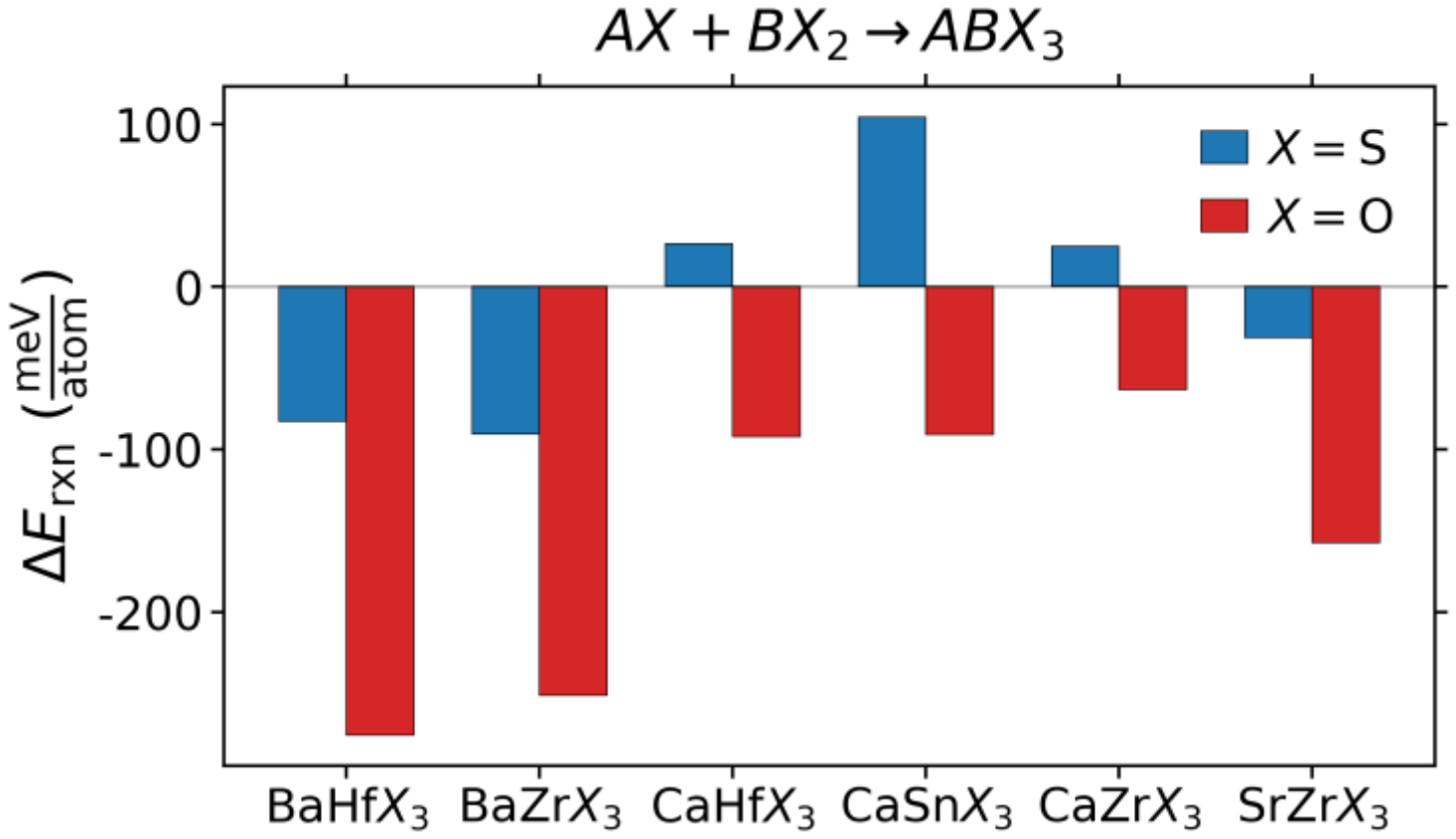

**Figure 6.** DFT reaction energies for $AX + BX_2 \rightarrow ABX_3$ reactions for 6 pairs of orthorhombic perovskites: $BaHfX_3$, $BaZrX_3$, $SrZrX_3$, $CaHfX_3$, $CaSnX_3$ and $CaZrX_3$ (where $X$ = O or S). The vertical position of each bar indicates its reaction energy ($\Delta E_{rxn}$), with more negative values indicating a larger driving force to form the $ABX_3$ compound.

One possible mechanism is the inductive effect, which has been explored in previous studies as a potential explanation for the stabilization of ternary compounds compared with their binary counterparts. The inductive effect can be understood as an electron pressure exerted when two electropositive cations are bonded to opposite sides of an anion in an *A*-*X*-*B* bonding environment.[66] If *A* is significantly more electropositive than *B*, it donates additional electron density to the anion (*X*), which in turn strengthens the *B*-*X* bond by making it more covalent. This effect has been documented in nitride systems (among others), such as CaNiN, where the introduction of a highly electropositive Ca cation stabilizes the compound relative to its unstable $Ni_3N$ binary.[67]

Ternary oxides likely benefit more from inductive stabilization than their sulfide counterparts due to the higher electronegativity of oxygen relative to sulfur. Previous studies have largely attributed the inductive effect to differences in cation electronegativity.[66,68,69] Extending this reasoning, one can infer that the strength of the inductive effect is also influenced by the electronegativity of the anion (*X*). We posit that an anion with lower electronegativity will be less effective at attracting electron density from the electropositive cation, thereby diminishing the stabilizing influence of the inductive effect. To test this hypothesis, in **Table 1**, we show the change in Bader charge for the *A*- and *B*-site cations ($\Delta\delta_A$ and $\Delta\delta_B$) calculated as the difference between the cation charge in the ternary compound ($ABX_3$) and in its corresponding binary ($AX$ or $BX_2$). We selected a subset of the compounds shown in **Figure 6** for this analysis. Separate values are presented for both sulfides and oxides. An indicator of the inductive effect is an increased charge ($\Delta\delta_A > 0$) of the *A*-site cation going from binary (e.g., CaO) to ternary (e.g., $CaHfO_3$) and a decreased charge of the $B$-site cation ($\Delta\delta_B < 0$) as the $A$-$X$ bonds become more ionic and the $B$-$X$ bonds become more covalent. Good examples to test this are $CaHfX_3$, $CaZrX_3$, and $CaSnX_3$. Both $CaHfX_3$ and $CaZrX_3$ have orthorhombic perovskite ground states as both oxides and sulfides, yet $CaHfS_3$ and $CaZrS_3$ exhibit slight hull instabilities ($\Delta E_d$ = 27 meV/atom and 25 meV/atom, respectively), while their oxide counterparts are hull-stable. $CaSnS_3$, in contrast, is both polymorphically and hull unstable in the perovskite structure ($\Delta E_{gs}$ = 44 meV/atom; $\Delta E_d$ = 61 meV/atom) whereas its oxide counterpart, $CaSnO_3$, is a stable perovskite.

**Table 1.** Bader charge differences and reaction energies of selected perovskite oxides and sulfides. The Bader charge difference for the $A$-site ($\Delta\delta_A$) and $B$-site ($\Delta\delta_B$) is calculated as the charge in the ternary chalcogenide minus that in the corresponding binary compound. For example, $\Delta\delta_A = \delta$(Ca in $CaHfS_3$) − $\delta$(Ca in CaS); $\Delta\delta_B = \delta$(Hf in $CaHfS_3$) − $\delta$(Hf in $HfS_2$). Also shown are the reaction energies ($\Delta E_{rxn}$, in meV/atom) for the formation of each ternary from its binary precursors (e.g., CaS + $HfS_2$ → $CaHfS_3$).

| | $\Delta\delta_A$ | $\Delta\delta_B$ | $\Delta E_{rxn}$ |
|---|---|---|---|
| CaS + $HfS_2$ → $CaHfS_3$ | 0.05 | 0.02 | 27 |
| CaO + $HfO_2$ → $CaHfO_3$ | 0.08 | -0.05 | -92 |
| CaS + $ZrS_2$ → $CaZrS_3$ | 0.05 | 0.02 | 25 |
| CaO + $ZrO_2$ → $CaZrO_3$ | 0.08 | -0.03 | -63 |
| CaS + $SnS_2$ → $CaSnS_3$ | 0.05 | 0.00 | 104 |
| CaO + $SnO_2$ → $CaSnO_3$ | 0.09 | -0.05 | -91 |

The data in **Table 1** reveals consistent trends in charge redistribution when going from binary to ternary compounds. In all systems, the $A$-site cation shows a larger increase in Bader charge when going from the binary to the ternary in oxides than in sulfides, indicating more electropositive behavior and stronger electron donation in oxides. Meanwhile, the $B$-site cation in oxides shows a clear decrease in charge from the binary to the ternary, consistent with increased covalency in the $B$–O bond due to the inductive effect. In contrast, sulfides show a smaller change or even an increase in $B$-site charge when forming the ternary. This contrasting behavior supports the notion of a weaker inductive effect in sulfides than oxides, potentially contributing to their differing thermodynamic stabilities and the relative sparsity of $ABS_3$ compared with $ABO_3$ compounds.

**Implications for selenide and telluride perovskites**

Although the broader class of chalcogenide perovskites includes selenides and tellurides, this study focuses on sulfide compounds due to their greater potential for optoelectronic applications. While a detailed analysis of selenides and tellurides is beyond the scope of this work, trends observed in our sulfide study allow us to speculate on their behavior. From a geometric perspective, the larger ionic radii of Se (1.98 Å) and Te (2.21 Å) compared to S (1.84 Å) decreases the number of $A$/$B$ cation combinations that fit the corner-sharing perovskite geometry. Of the 81 *A*/*B* combinations analyzed, 17 $ABS_3$ compounds fall within the perovskite-stable region of the Jess tolerance factor whereas only 6 $ABSe_3$ and 5 $ABTe_3$ materials fall within that region. These selenide and telluride counts are based on a slightly narrower Jess tolerance factor range (1.01–1.12),[35] which was defined using known selenide perovskites. In the absence of known tellurides, the same range was applied for $ABSe_3$ and $ABTe_3$ compounds.

According to the Turnley tolerance factor, initial filtering using the μ criterion allows 45 sulfides, 27 selenides, and 0 tellurides to pass, implying a substantially smaller pool of selenides and tellurides capable of forming $BSe_6$ or $BTe_6$ octahedra. Electronegativity trends further support the sparsity of telluride perovskites. In **Figure 3b**, we observe that instability with respect to other polymorphs ($\Delta E_{gs}$) grows with decreasing $\chi_{diff}$. Se is only slightly less electronegative than S (2.55 vs. 2.58), but Te is significantly less (2.10), placing all telluride compositions below the $\chi_{diff}$ threshold of 1.025 defined by the Turnley tolerance factor. As a result, no $ABTe_3$ compounds and only 6 $ABSe_3$ compounds satisfy the Turnley conditions for perovskite stability, compared to 9 $ABS_3$ compounds.

In our polymorphic instability analysis, we found that increased covalency in *B*–*X* bonding enhances the energetic stability of the perovskite relative to the needle structure. Because selenides have similar electronegativities to sulfides, they likely exhibit comparable covalent stabilization. Tellurides, on the other hand, may experience stronger *B*–*X* bonding interactions, which could theoretically favor perovskite formation by reducing competition with the needle structure. While increased covalency might promote polymorphic stability, the larger ionic radius of Te makes it more difficult to form stable $BX_6$ octahedra, as discussed previously.

In terms of thermodynamic stability with respect to phase separation, sulfide phase diagrams are the most populated with 196 stable competing compounds in the 81 *A*-*B*-S chemical spaces. There are 170 and 148 stable competing compounds in the corresponding *A*-*B*-Se and *A*-*B*-Te chemical spaces, respectively. This may reflect either weaker competition for ternary compound formation or indicate a relative lack of exploration within the Materials Project database for these less common anions. Moreover, as we discussed in our hull stability analysis, a decrease in electronegativity leads to diminished inductive stabilization. This could help explain the lack of experimental synthesis reports of $ABTe_3$ compounds. In total, our analysis suggests that $ABSe_3$ perovskites face comparable stability challenges as $ABS_3$ perovskites, while $ABTe_3$ perovskites are significantly less likely to form.

**Conclusion**

This study probes the thermodynamic stability of chalcogenide perovskites and reveals that very few are thermodynamically stable. By assessing these instabilities with respect to polymorphism and decomposition into competing compounds, we identify chemical and thermodynamic factors governing the scarcity of stable chalcogenide perovskites. In terms of polymorphism, the needle structure was calculated to be the ground-state polymorph for 77% of the 81 compounds analyzed in this study. Of the few compounds with perovskite ground states, covalent *B*-S bonding was found to be a significant driver for stability. In terms of stability with respect to decomposition, we propose that a weak inductive effect for ternary sulfide formation leads to lower thermodynamic driving force for $ABS_3$ formation compared to analogous $ABO_3$ compounds. Our findings demonstrate that recently proposed tolerance factors adapted for chalcogenide perovskites tend to overpredict perovskite stability as we identify several hypothetical $ABS_3$ that are unstable in the perovskite structure yet lie within the bounds of these tolerance factors where perovskites are expected. This underscores that accurate prediction of chalcogenide perovskite stability requires consideration of multiple factors beyond geometry. While few $ABS_3$ perovskites are calculated to be thermodynamically stable, several hypothetical materials are computed to be weakly unstable and therefore plausibly accessible under carefully chosen synthetic conditions. In total, this work leads to new understanding of the thermodynamics and crystal chemistry of this compelling class of materials.

## Acknowledgments

This work was supported by the National Science Foundation Division for Materials Research Award No. 2433203. The authors acknowledge the Minnesota Supercomputing Institute (MSI) at the University of Minnesota for providing resources that contributed to the research results reported herein. The authors also acknowledge Lauren Borgia and Yi-Ting Cheng for helpful discussions during the development of this work.

**Supplementary Information**

**Origins of chalcogenide perovskite instability**

Adelina Carr[a], Talia Glinberg[b], Nathan Stull[c], James R. Neilson[c,d], Christopher J. Bartel[a*]

[a] University of Minnesota, Department of Chemical Engineering and Materials Science, Minneapolis, MN 55455

[b] University of Minnesota, Department of Chemistry, Minneapolis, MN 55455

[c] Colorado State University, Department of Chemistry, Fort Collins, CO 80523

[d] Colorado State University, School of Materials Science & Engineering, Fort Collins, CO 80523

* correspondence to cbartel@umn.edu

**Table S1.** DFT-calculated polymorphic stability, DFT-calculated decomposition energies, and tolerance factor predictions for $ABS_3$ compounds. The full dataset is provided in the accompanying file **table_S1.csv**. Radii used for the Jess tolerance factor[1] are Shannon ionic radii derived from oxides and halides[2]; for the Turnley tolerance factor,[3] Shannon radii derived from sulfides[4] are used. The sulfur ionic radius is 1.84 Å for the Jess factor and 1.70 Å for the Turnley factor.

Column definitions:

- **Compound:** $ABS_3$ compound formula
- **A / B:** *A*- and *B*-site elements
- **synthesized:** True if compound has been previously synthesized in any structure
- **calc_gs:** DFT-calculated ground state structure
- **$\Delta E_{gs}$:** DFT-predicted energy difference between the perovskite structure and the corresponding ground-state structure
- **$\Delta E_d$:** DFT predicted decomposition energy of the ground-state polymorph relative to the convex hull
- **Predicted_decomp_products:** Predicted products from the decomposition reaction that defines thermodynamic stability relative to the convex hul
- **Number_of_competing_phases:** Number of thermodynamically stable competing compositions in the respective *A*-*B*-S phase diagram, reported as: total number of compositions (unary, binary, ternary), based on their compositional complexity
- **tf_Jess ($t^*$):** Jess tolerance factor value
- **tf_Jess_pred:** predicted structure based on the Jess tolerance factor
- **rA_Jess / rB_Jess:** *A*/*B*-site radii used for the Jess tolerance factor
- **tf_Turnley (t^S):** sulfide-adjusted Goldschmidt tolerance factor value used in the Turnley model
- **tf_Turnley_pred:** predicted structure based on the Turnley model
- **$\chi_{diff}$ :** electronegativity difference used in the Turnley model, $\chi_{diff} = 1/5\ (3\chi_S - \chi_A - \chi_B)$
- **μ:** octahedral factor value used in the Turnley model, $\mu = r_B / r_X$
- **rA_Turnley / rB_Turnley:** *A*/*B*-site radii used for the Turnley model
- **DOI:** reference DOI for experimentally reported compounds

**Table S2.** Decomposition energies relative to the convex hull for the 81 $ABS_3$ compounds analyzed and their 81 $ABO_3$ oxide analogues. The full dataset is provided in the accompanying file **table_S2.csv**. Sulfide decomposition energies are calculated in this work,

while oxide decomposition energies are taken from the Materials Project database[5] where available. Oxide entries that are not present in Materials Project are left blank.

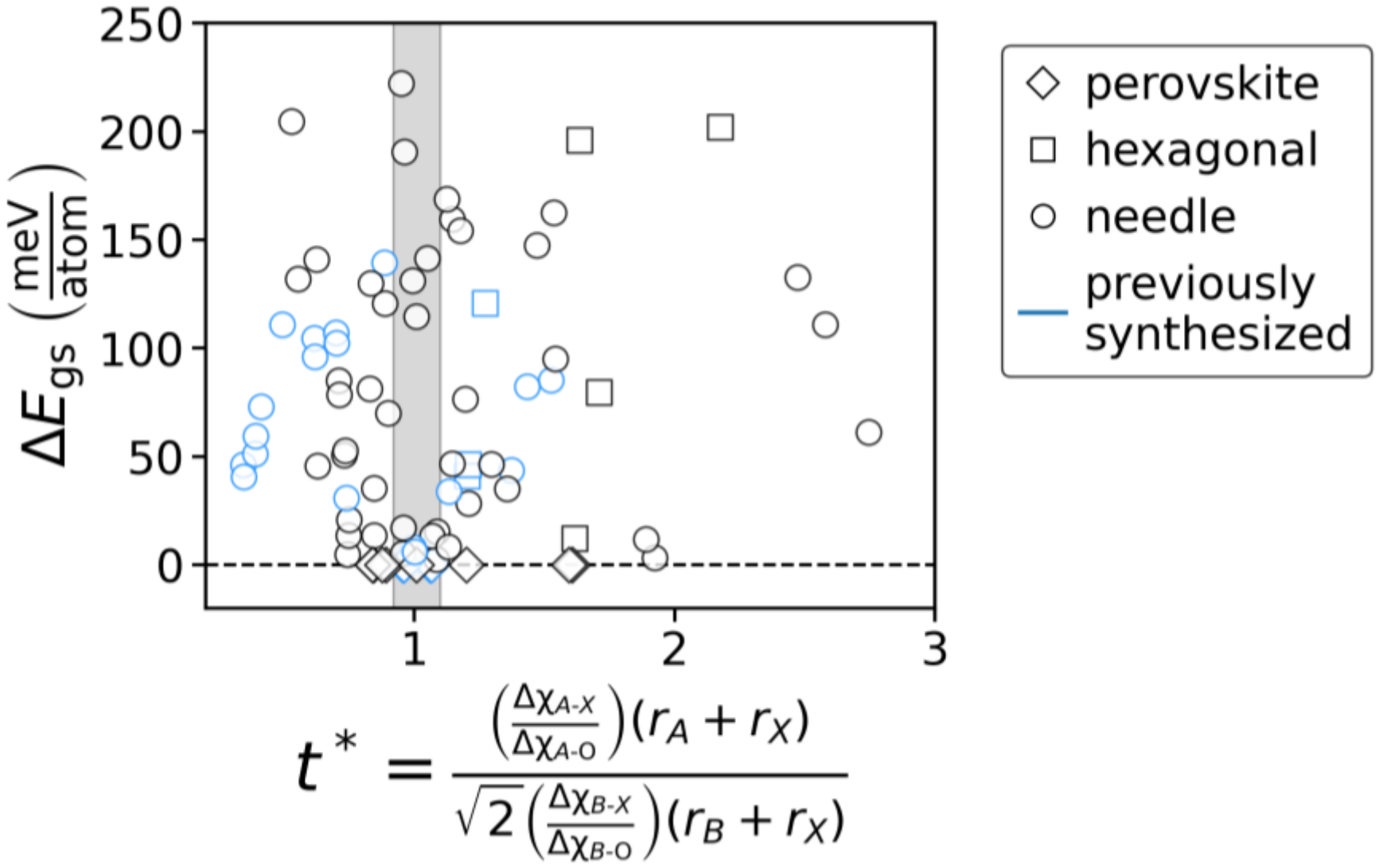

**Figure S1.** Expanded view of **Figure 3a**. DFT-predicted energy difference between the perovskite structure and the corresponding ground-state structure ($\Delta E_{gs}$) plotted against the Jess tolerance factor ($t^*$) for all 81 materials analyzed, where points at $\Delta E_{gs} = 0$ indicate perovskite stability. The gray region denotes the $t^*$ range where this tolerance factor predicts perovskite stability. Blue-outlined data points denote previously synthesized compounds (in any structure), and marker shapes indicate the DFT-predicted ground-state structure: diamonds for perovskite, circles for needle, and squares for hexagonal.

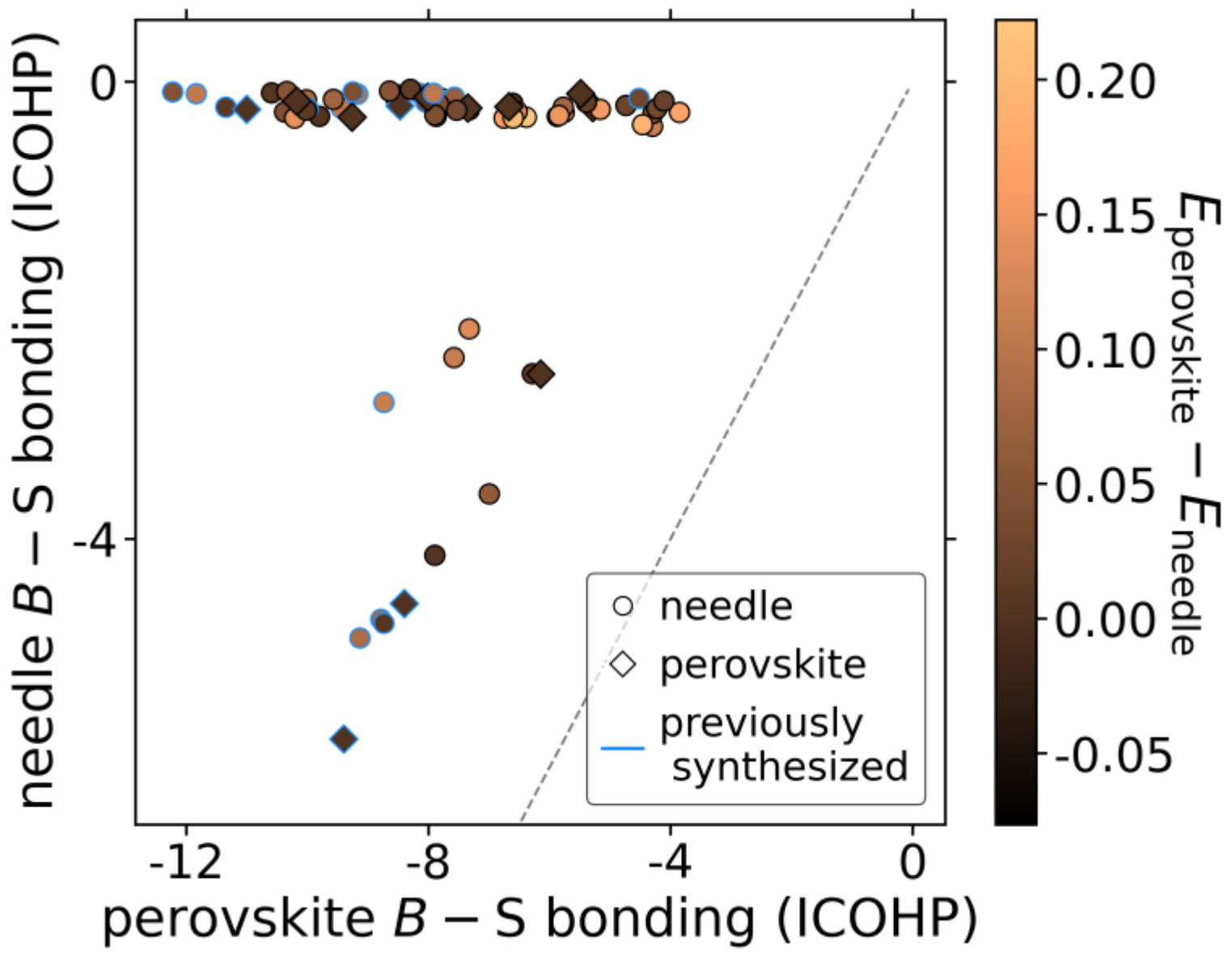

**Figure S2.** Parity plot comparing integrated COHP (ICOHP) values for *B*–S bonding interactions in the needle and perovskite structures, normalized per formula unit and integrated from 0 to –7 eV below the Fermi level. Negative values represent bonding interactions, while positive values represent antibonding interactions. The dashed line indicates parity (equal *B*–S bonding strength in both structures). Points are colored according to the difference in energy from the perovskite to the needle structures. Points outlined in blue correspond to previously synthesized materials. Diamond-shaped markers represent materials that have a perovskite ground state, while circles represent a needle ground state (according to our calculations). Because we compare needle and perovskite only, we have excluded materials with a hexagonal ground state. Here, we have only considered materials with differing *A*- and *B*-site atoms. The units for ICOHP are arbitrary.

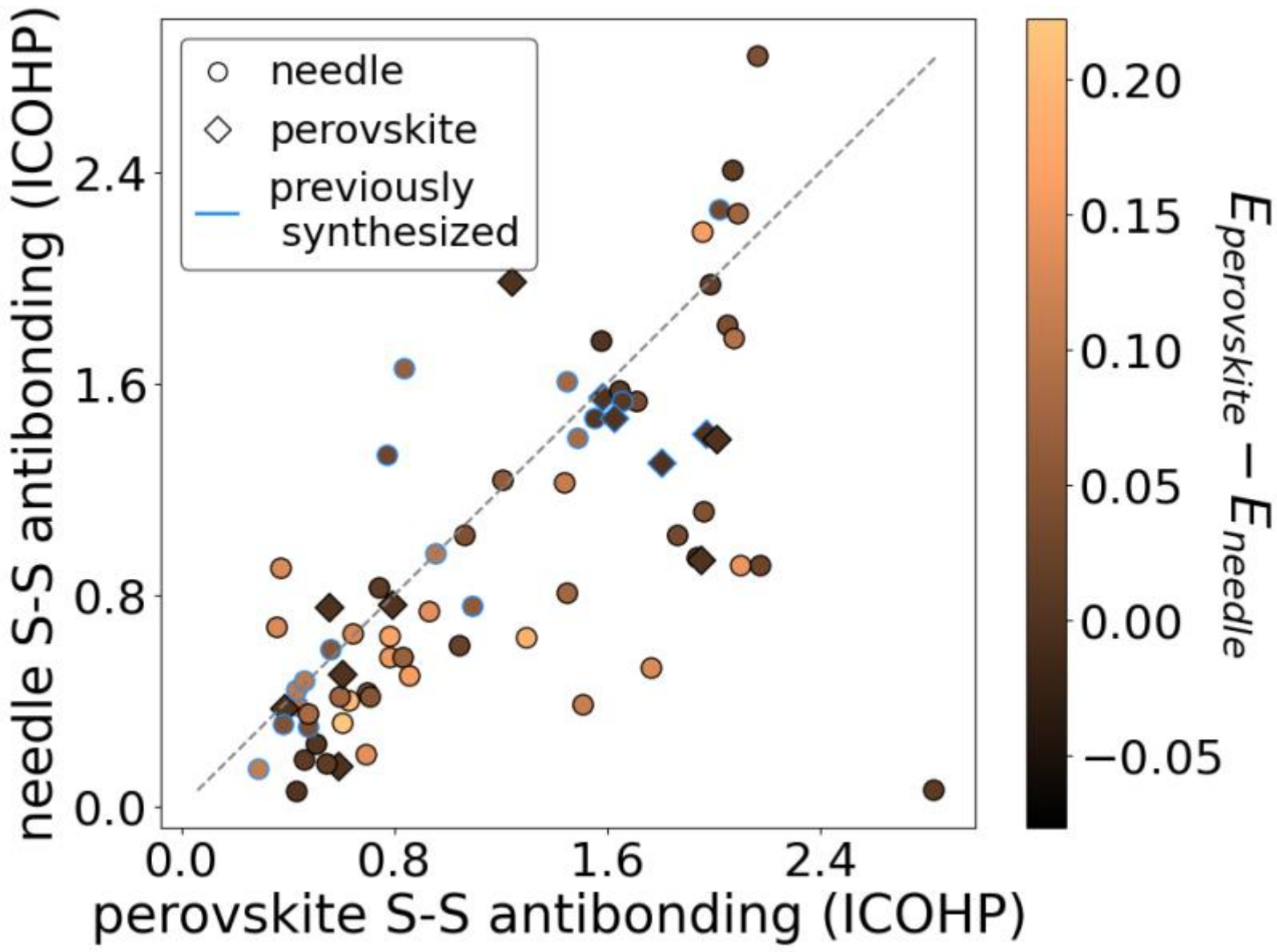

**Figure S3.** Parity plot comparing integrated COHP (ICOHP) values for S–S antibonding interactions in the needle and perovskite structures, normalized per formula unit and integrated from 0 to –7 eV below the Fermi level. Negative values represent bonding interactions, while positive values represent antibonding interactions. Points are colored according to the difference in energy from the perovskite to the needle structures. Points outlined in blue correspond to previously synthesized materials. Diamond-shaped markers represent materials that have a perovskite ground state, while circles represent a needle ground state (according to our calculations). Because we compare needle and perovskite only, we have excluded materials with a hexagonal ground state. Here, we have only considered materials with differing *A*- and *B*-site atoms. The units for ICOHP are arbitrary.

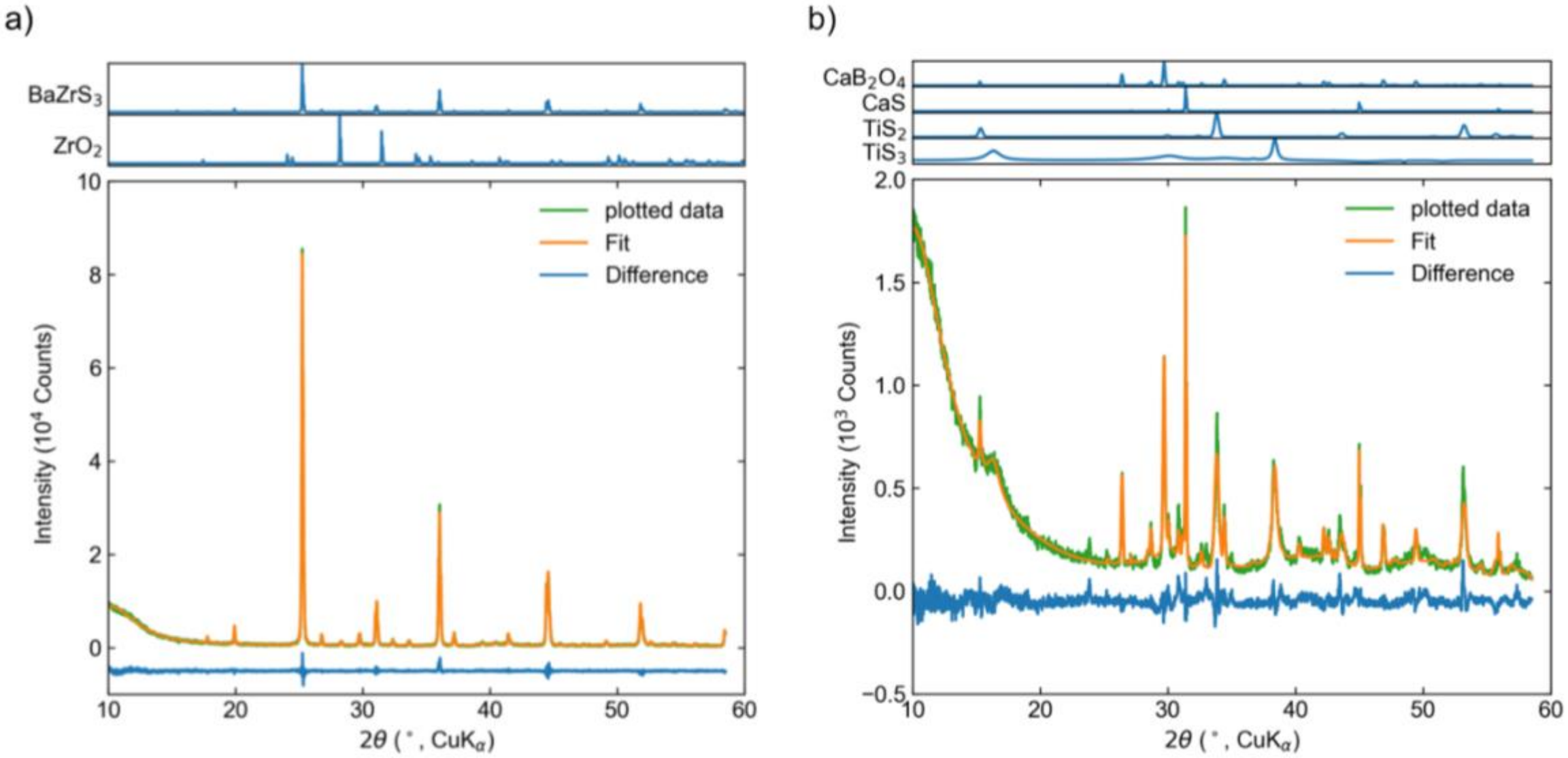

**Figure S4.** PXRD data for the attempted synthesis of a) $BaZrS_3$ and b) $CaTiS_3$. PXRD data showed a pure phase sample of $BaZrS_3$. Analysis of the PXRD data of the reaction product targeting $CaTiS_3$ showed multiple metal sulfide (CaS, $TiS_2$, and $TiS_3$) products as well as $CaB_2O_4$. The reaction did not yield any Ca-Ti-S ternary phases.